\documentclass[aps,preprint,nofootinbib]{revtex4}
\usepackage{graphicx}
\usepackage{color}
\usepackage{epsfig}

\begin{document}
\title{Strong Correlations and Fickian Water Diffusion  
in Narrow Carbon Nanotubes}

\author{Biswaroop Mukherjee$^{1,3,}$}
\email{biswa@physics.iisc.ernet.in}
\author{Prabal K.~Maiti$^{1,}$}
%\thanks{To whom correspondence should be addressed}
\email{maiti@physics.iisc.ernet.in}
\author{Chandan Dasgupta$^{1,3,}$}
\email{cdgupta@physics.iisc.ernet.in}
\author{A.~K.~Sood$^{2,}$}
\email{asood@physics.iisc.ernet.in}
\affiliation{
$^1$Centre for Condensed Matter Theory, Department of Physics, Indian Institute of Science,
Bangalore 560 012, India \\
$^2$ Department of Physics, Indian Institute of Science, Bangalore 560 012, India \\
$^3$ Condensed Matter Theory Unit,
Jawaharlal Nehru Centre for Advanced Scientific Research, Bangalore 560 064, India.}

\begin{abstract}
We have used atomistic molecular dynamics (MD) simulations to study the structure 
and dynamics of water molecules inside an open ended carbon nanotube placed in a 
bath of water molecules. The size of the nanotube allows only a single file of 
water molecules inside the nanotube.
The water molecules inside the nanotube show solid-like ordering at 
room temperature, which we quantify by calculating the pair correlation function.  
It is shown that even for the longest observation times, the
mode of diffusion of the water molecules inside the nanotube 
is Fickian and not sub-diffusive. We also
propose a one-dimensional random walk model for the diffusion of the
water molecules inside the nanotube. We find good agreement
between the mean-square displacements calculated from the random walk model and 
from MD simulations, thereby confirming that the water molecules undergo
normal-mode diffusion inside the nanotube. We attribute this behavior to strong
positional correlations that cause all the water molecules inside the nanotube
to move collectively as a single object. The average residence time 
of the water molecules inside the nanotube is shown to scale
quadratically with the nanotube length.
\end{abstract}

\maketitle

\section{Introduction}
Fluids confined in nanometric scales are known to behave very differently from
their bulk counterpart \cite{zangi}. Structurally, they show solid-like ordering, 
induced by the confinement, whereas their diffusion constants 
are of the same order of magnitude as those of ordinary bulk liquids. 
Water confined in carbon nanotubes serves as an
ideal example of such a system. The interest in studying such systems 
arises from the importance of understanding the properties of
confined water and from the possibility of extrapolating the
conclusions to other physical situations of similar nature, such as water
adsorbed in nanopores in biological and geological systems. 
In recent years, Molecular Dynamics (MD) simulations have been used 
to study various aspects of these systems in microscopic detail
\cite{hummer_nature,beckstein1,beckstein2,Koga,Koga_jcp,Striolo1,Striolo_single}.
Such studies complement 
experimental investigations by providing detailed microscopic understanding of
some of the experimentally observed features. In some cases, simulations also
make interesting predictions that can be tested in experiments.

Both structural and dynamical properties of water molecules in nanotubes have been
investigated in MD simulations.
Water molecules confined in nanotubes exhibit a variety of  
spatially ordered structures not seen in bulk water. MD studies of Koga {\it et al.}
\cite{Koga,Koga_jcp} predict  the occurrence of hexagonal,
heptagonal, square or pentagonal ice nanotube structures
inside carbon nanotubes of diameter ranging from $11$ to $14 \AA$ under different
conditions. Some of these structures have been inferred from  
x-ray diffraction \cite{Maniwa}  and neutron scattering \cite{Kolesnikov}
experiments. Striolo {\it et al.} \cite{Striolo1} have performed
Grand Canonical Monte Carlo (GCMC) simulations to study the adsorption of water in
single-walled carbon nanotubes and found structures
similar to those observed in neutron scattering experiments
\cite{Kolesnikov}. GCMC simulations on adsorption of organic mixtures of 
cyclohexane, methane, ethane etc. inside carbon nanotubes show ordered arrangement
of the adsorbed molecules \cite{ayappa}.

Among the dynamical properties of confined water, one that has attracted 
a lot of attention is  the process of conduction of water molecules
through nanopores. It is believed that this process is important in
understanding the transport through biological ion channels \cite{Tajkhorshid}.
Hummer {\it et al.} \cite{hummer_nature} studied the conduction of water molecules 
through narrow carbon nanotubes using MD simulations. They found that clusters of 
water molecules are transferred through the nanotube in the form of 
occasional bursts arising due to pressure fluctuations occurring 
outside the nanotube. Beckstein {\it et al.} \cite{beckstein1,beckstein2}  
have extended such simulations to study the 
density fluctuations of water inside other hydrophobic pores.
Transport of oxygen and organic molecules such as
methane, ethane and ethylene through carbon nanotubes has also been
studied \cite{Lee1,Mao,Lee2,yashonath} through MD simulations. 

Another interesting question about the dynamics of water molecules in narrow
carbon nanotubes is concerned with the mode of diffusion of the water 
molecules. If the diameter of the nanotube is comparable to the size of a water
molecule, then the water molecules can not cross one another while
they are inside the nanotube and they have to move in a single file. 
Various earlier studies \cite{harris,jespen,beijeren,alexander,richards,fedders,
levitt1,karger1,karger2,hahn,bagchi} 
suggest that under these conditions, 
the water molecules inside the nanotube should exhibit a kind of
sub-diffusive behavior known as {\it single file diffusion} 
where the mean-square displacement (MSD) scales with time $t$ as $t^{1/2}$
at long times. A crossover from normal mode diffusion (MSD $\propto t$) to 
single file behavior has been observed in experiments on colloidal particles
suspended in water inside narrow trenches \cite{lin1,wei,lutz,lin2} and on 
molecules diffusing through channels in zeolites \cite{zeolite}.

The question whether molecules exhibit single file diffusion inside narrow
carbon nanotubes has been investigated in a number of simulation studies.
Lee {\it et al.} \cite{Lee1} have studied
the diffusion of oxygen molecules inside
open-ended carbon nanotubes of various diameters. They interpret
their simulation results implying that oxygen
diffuses through wide (17,0) nanotubes via normal mode diffusion, whereas
inside narrow (10,0) nanotubes, the molecules exhibit sub-diffusive
behavior (MSD  $\sim t^{\alpha}$, where
the exponent $\alpha$ is less than unity). 
The simulations of Mao {\it et al.} \cite{Mao} suggest that
spherical methane molecules exhibit normal mode  diffusion inside closed
carbon nanotubes, whereas aspherical molecules such as ethane and ethylene 
diffuse by a mode which is between
normal and single-file. In all these simulations, the time over which the MSD 
is measured is rather short ($100-500$ ps). Also, due to finite-size effects
arising from the smallness of the nanotube length (these effects will be
discussed in detail below), it is difficult to draw a firm conclusion about the 
true nature of diffusion inside the nanotube from the time-dependence of the
MSD. For these reasons, we believe that the simulations mentioned above do
not provide conclusive evidence for sub-diffusive behavior of molecules inside
a carbon nanotube. Recently, Striolo \cite{Striolo_single} has performed
MD simulations of the diffusion of water molecules inside a (8,8)
nanotube using periodic boundary conditions in the
direction of the axis of the nanotube. This artificial boundary condition
prevents the water molecules from escaping the nanotube and hence
it is possible to obtain MSD data at long times with good statistics.
These simulations show that the water molecules inside the nanotube 
initially undergoes ballistic diffusion (MSD $\propto t^2$)
which crosses over to normal mode diffusion at long times. 

In this work we use MD simulations to study the structural and
dynamical properties of
water molecules inside (6,6) carbon nanotubes of diameter $8 \AA$
and lengths $L$ equal to $
14$, $28$ and $56 \AA$ in a bath of water. 
Since the water
molecules inside the nanotube can not cross one another, 
a primary goal of our study is
to determine whether single file diffusion occurs in this system. 
Here the interactions among the water molecules 
inside the nanotube are
much stronger compared to those among the colloidal particles considered in
experiments \cite{lin1,wei,lutz,lin2} that demonstrate the 
occurrence of single file diffusion.
This can be inferred from the packing fraction
of the water molecules inside the tube. The packing fraction, $\eta$, is
defined as $\eta = N \sigma/L$, where $N$ is the number of water
molecules, $\sigma$ the diameter of each molecule and $L$, the length of
the tube. This quantity is close to unity for the water molecules inside
the (6,6) nanotube (it depends to some extent on the equilibrium density 
of the water molecules in the bath). This introduces strong correlations
in the positions and displacements of the water molecules in the nanotube.
These correlations and the effects of the open ends of the relatively short
nanotubes considered in our study (in particular, the fact that the range of
correlations is comparable to the length of the nanotube) have important
bearings on the  diffusion behavior of the water molecules.  

We find that the confined 
water molecules are spatially ordered, much  
like a solid, at temperatures as high as $300$ K. 
The MSD of the confined water molecules increases linearly with 
time $t$ at relatively short times (after a very short
initial regime of quadratic increase characteristic of ballistic motion),
indicating normal-mode diffusion over this time scale. At long times the 
MSD eventually saturates. We show, by comparing the
simulation results with analytic ones obtained from a simplified random walk model
described below, that the observed behavior is completely consistent
with normal mode diffusion of the water molecules inside the nanotube. In
particular, we observe that the departure of the MSD versus $t$ curve from the
initial linear behavior is a consequence of biased sampling arising from the
fact that the MSD is measured only for the molecules that stay inside the nanotube.
This conclusion is
consistent with existing results \cite{hummer_prl} that show that the statistics
of the bursts of water molecules passing through a narrow carbon nanotube
\cite{hummer_nature} are consistent with one-dimensional random walk behavior.
Our work also brings out that the normal mode diffusion found recently by
Striolo \cite{Striolo_single} for water molecules in narrow carbon nanotubes with
periodic boundary conditions persists in the more realistic situation of
open-ended nanotubes immersed in water.

The remainder of the paper is organized as follows.
Section 2 describes the details of the simulation. In Section 3 we
describe the results obtained from the MD simulations for the static 
structure and the dynamics of the water molecules inside the 
nanotube. In Section 4 we propose a random walk model to explain the 
dynamics of the confined water molecules and compare the analytic results
for the model with those obtained from our MD simulations.
Section 5 contains a summary of our main results and a few concluding remarks.

\section{Details of Simulation}
We have simulated
a (6,6) carbon nanotube of $8\AA$ diameter placed in a bath of
TIP3P \cite{jorgensen} water molecules at $300$K
and $1$ atmosphere pressure. During the simulation the nanotube was
held fixed inside the simulation box, with its long axis along the
$z$ direction.
The simulations have been performed using AMBER 7 \cite{amber}.
We have considered open-ended nanotubes of length $L$ equal to $14 \AA$,
$28 \AA$ and $56 \AA$. A given simulation contained 
1000 to 2200 water
molecules depending on the size of the nanotube used.
The interactions between various atoms have been described by
classical force fields. The carbon and oxygen atoms were modelled
as Lennard-Jones particles, the details of which are given in
Table I. 
%\ref{lenjonesparm}.
The carbon-carbon bond length was $1.4 \AA$ and the 
corresponding spring constant
was $938$ kcal/(mol $\AA^{2})$, the equilibrium C-C-C angle 
was $2\pi/3$ radians
and the corresponding spring constant was $126$ kcal/(mol $rad^{2})$.
The simulations were performed for periods varying between 8 ns
and $20$ ns (longer simulations were carried out for nanotubes with larger
$L$) and the coordinates were stored at
an interval of $1$ ps.

\section{Results}
To quantify
the positional ordering of the water molecules inside the nanotube we have
calculated the pair-correlation function for the water molecules inside the nanotube
using
\begin{eqnarray}
g(z) = \frac{1}{N} \sum_{i=1}^N \sum_{j=1,j\ne i}^N <\delta(z - z_{ij})>,
\label{gz}
\end{eqnarray}
where $z_{ij}$ is the axial separation between the $i$ th and the $j$ th water molecules,
N is the number of water molecules inside the nanotube and the angular
brackets indicate an average over time.

\begin{figure}[htbp]
\includegraphics[height=8.5cm,width=8.5cm,angle=0]{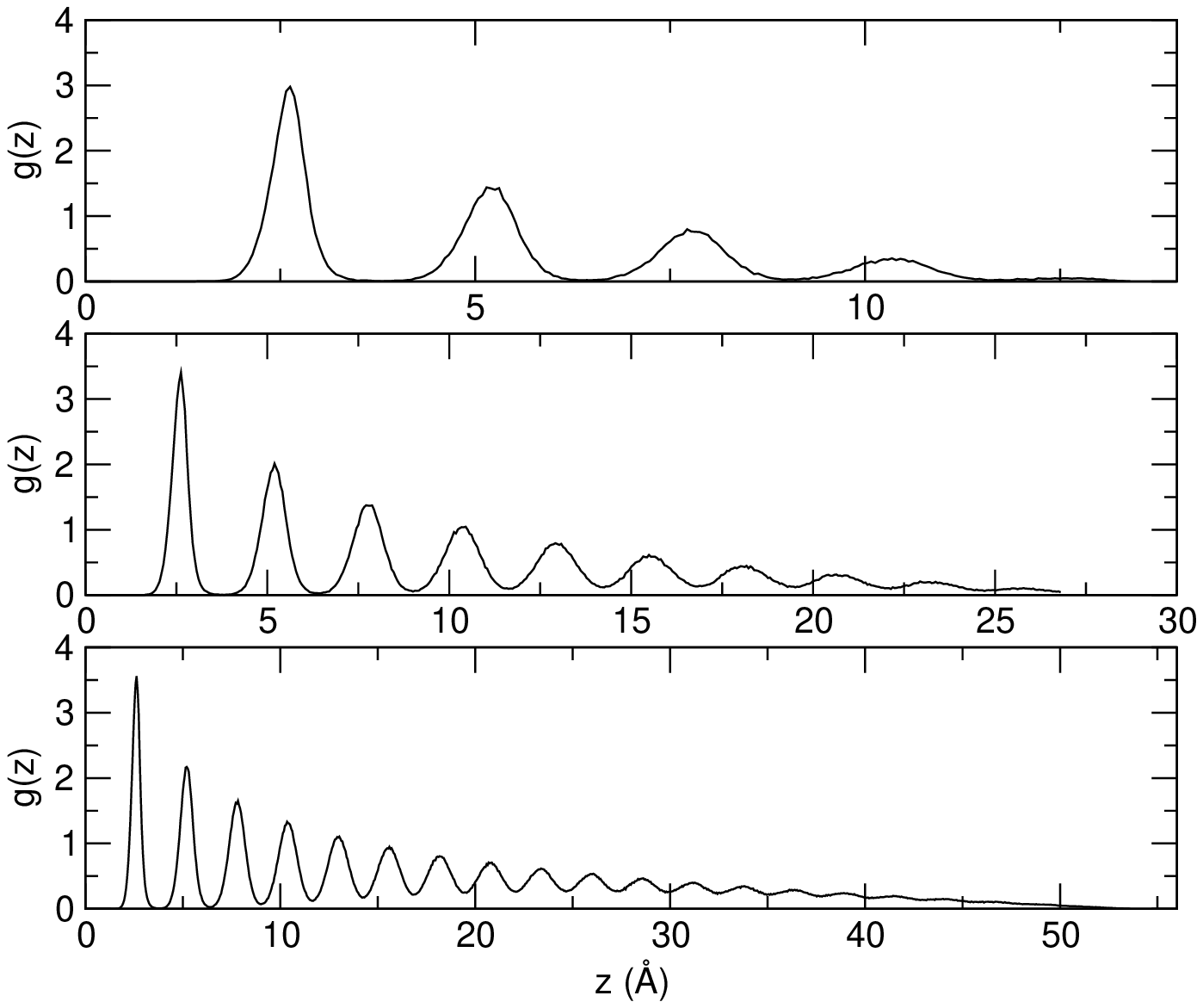} \\
\caption{}
\label{802709JCP1}
\end{figure}

\begin{figure}[htbp]
\includegraphics[height=8.5cm,width=8.5cm,angle=0]{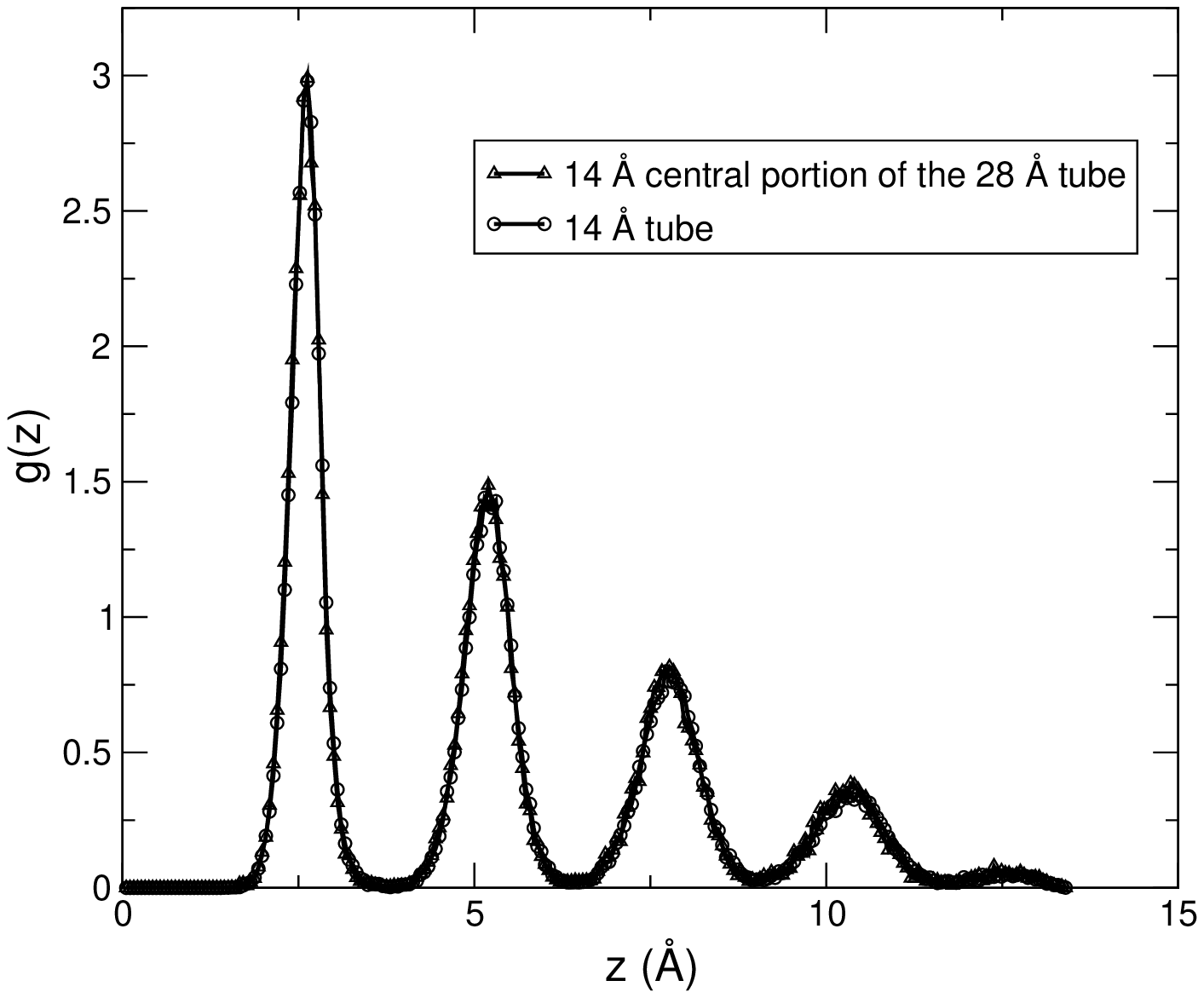} \\
\caption{}
\label{802709JCP2}
\end{figure}

The calculated pair correlation functions for nanotubes of length 
$14 \AA$, $28 \AA$ and
$56 \AA$ are shown in Fig. \ref{802709JCP1}.
The distinct peaks of $g(z)$ suggest that there is solid-like ordering
of the water molecules inside nanotube, even at a temperature of  $300 K$,
for all the three tube lengths. The nearest-neighbour distance
between the water molecules is $2.6 \AA$, similar to
that obtained by Hummer {\it et al.} \cite{hummer_prl}.
The heights of the peaks of $g(z)$ are smaller for shorter nanotubes and the
peak height decreases faster with $z$ as the length of the nanotube is
reduced. However, this does not imply that the degree of positional order
of the water molecules inside the nanotube 
decreases as the tube length is reduced.
The differences between the results for $g(z)$ for nanotubes of different
length arise from a simple finite-size effect, namely that the
fraction of molecules that have a smaller number of neighbors due to their
proximity to the open ends of the nanotube is larger for smaller tubes.
This finite-size effect on the form of $g(z)$ is seen clearly on comparing 
the $g(z)$ obtained for the $14 \AA$ tube with that calculated for the 
water molecules residing in the
central $14 \AA$ portion of the $28 \AA$ nanotube, taking only the molecules
in this central region in the sums in Eq.(\ref{gz}). This comparison is
shown in Fig.\ref{802709JCP2}. The observation that the two correlation
functions are identical confirms that the degree of positional
correlations is the same in nanotubes of length $14 \AA$ and $28 \AA$. 
That the water molecules are correlated over distances comparable to the
length of the nanotube has important consequences on the diffusion
behavior of the molecules, as discussed below.

\begin{figure}[htbp]
\includegraphics[height=8.5cm,width=8.5cm,angle=0]{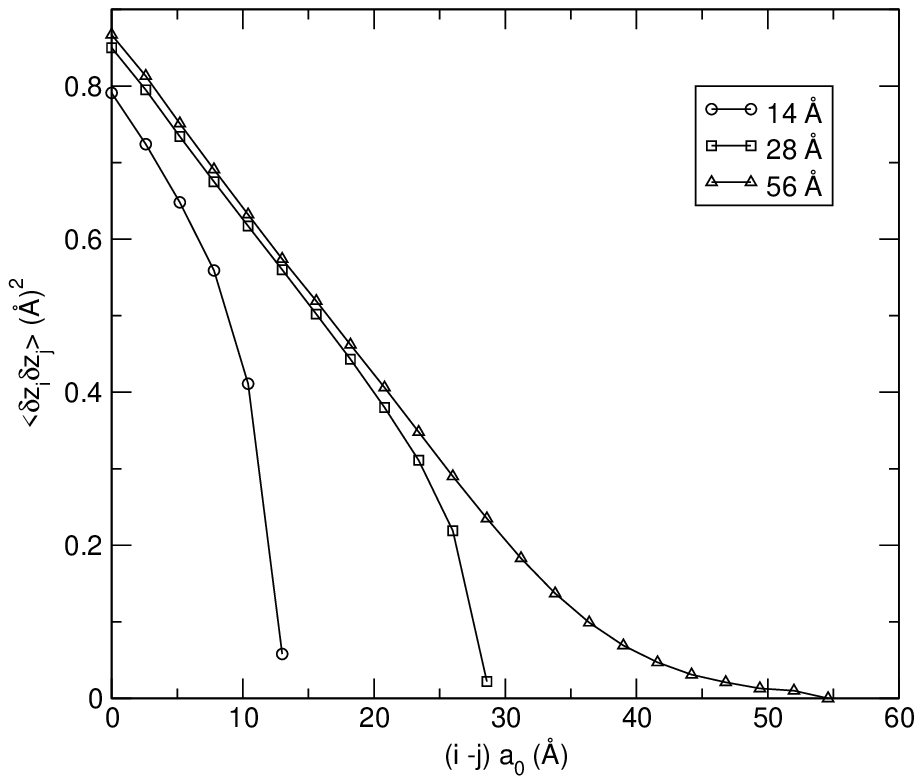} \\
\caption{}
\label{802709JCP3}
\end{figure}

We find that the trajectories of neighbouring molecules inside the nanotube
are extremely correlated. 
In order to quantify this correlation we have calculated the quantity,
$< \delta z_{i}(\delta t) \delta z_{j}(\delta t) >$, 
where $\delta z_{i}(\delta t)$  and $\delta z_{j}(\delta t)$ are
the displacements of the $i$th and $j$th water molecules, respectively, 
in a time interval of $\delta t$.
For this calculation, $\delta t$ was taken to be $1$ ps. 
The angular brackets indicate
averaging over time origins and pairs of water molecules inside the nanotube. 
This quantity
was calculated as function of the separation $z$ between the molecules
$i$ and $j$ for nanotubes of all three lengths. Results for 
this correlation function
are shown in Fig. \ref{802709JCP3}.
It decays as the separation $z$ is increased, 
signifying the loss
of correlation of the displacements of widely separated particles.
The separation at which this correlation function
decays to zero is close to the length of the tube, implying that 
the displacements of all the
molecules inside the tube are correlated for tubes of all three lengths. This
observation, consistent with the positional correlations shown in Fig.
\ref{802709JCP1}, implies that the water molecules inside the
nanotube tend to move together as a single object. 

The MSD of the water molecules inside the nanotube was
calculated using
\begin{eqnarray}
<\Delta z^{2} (t)> = \frac{1}{N} \sum_{i = 1}^{N} <[z_{i}(t + t') - z_{i}(t')]^{2}>_{t'},
\label{msd}
\end{eqnarray}
where $t$ is the time difference, $t'$ is a time-origin and
$N$ is the number of water molecules.
The angular brackets indicate an average over 
time origins.
The open ends of the tube allow particle exchange with the water bath.
As a result, while calculating the quantity,
$[z_{i}(t + t') - z_{i}(t')]^{2}$, for the $i$th water molecule, one has to
ensure that this molecule has stayed within the nanotube 
for the entire interval
of time between $t'$ and $(t + t')$. To ensure this, we have multiplied
$[z_{i}(t + t') - z_{i}(t')]^{2}$ by $\prod_{t'}^{(t+t')} P_{i}(t'')$,
where the
quantity $P_{i}(t)$ is equal to unity if the $i$th molecule is inside the
nanotube at time $t$ and zero otherwise. Similarly the number of time-origins
which contribute
for a particular molecule $i$ and time difference $t$ is
$\sum_{t'} \prod_{t'}^{(t+t')} P_{i}(t'')$. This way, even though the sum
in Eq.(\ref{msd}) goes over
all the water molecules present in the system, only those
molecules that have entered the tube at least once contribute in the
calculation of the MSD.

The finite lifetime of the water molecules inside the nanotube
makes it difficult to find sufficient number of long-lived water molecules,
the identification of which is crucial for good statistics of the
MSD at long times.
Hence, the simulation was run for $8$ ns for the system with $14 \AA$
nanotube to get MSD data up to $200$ ps. The run-time of the system
with $28 \AA$ nanotube was $12$ ns to get MSD data for a period of $1$ ns.
For the $56 \AA$ nanotube, even after running the simulation for about
$16$ ns, we were unable to get good quality data for the
MSD at long times. In this method, the short time data are obtained from averages 
over a large number of molecules, about 300 to 400, and also
a large number of time origins. The number of time-origins as well
as the number of water molecules available for averaging decrease as
one goes to longer time difference $t$. For this reason, the long-time MSD data
are more noisy. At the longest time
differences for which the MSD has been reported in this work, the
number of molecules contributing is approximately 50.

\begin{figure}[htbp]
\includegraphics[height=8.5cm,width=8.5cm,angle=0]{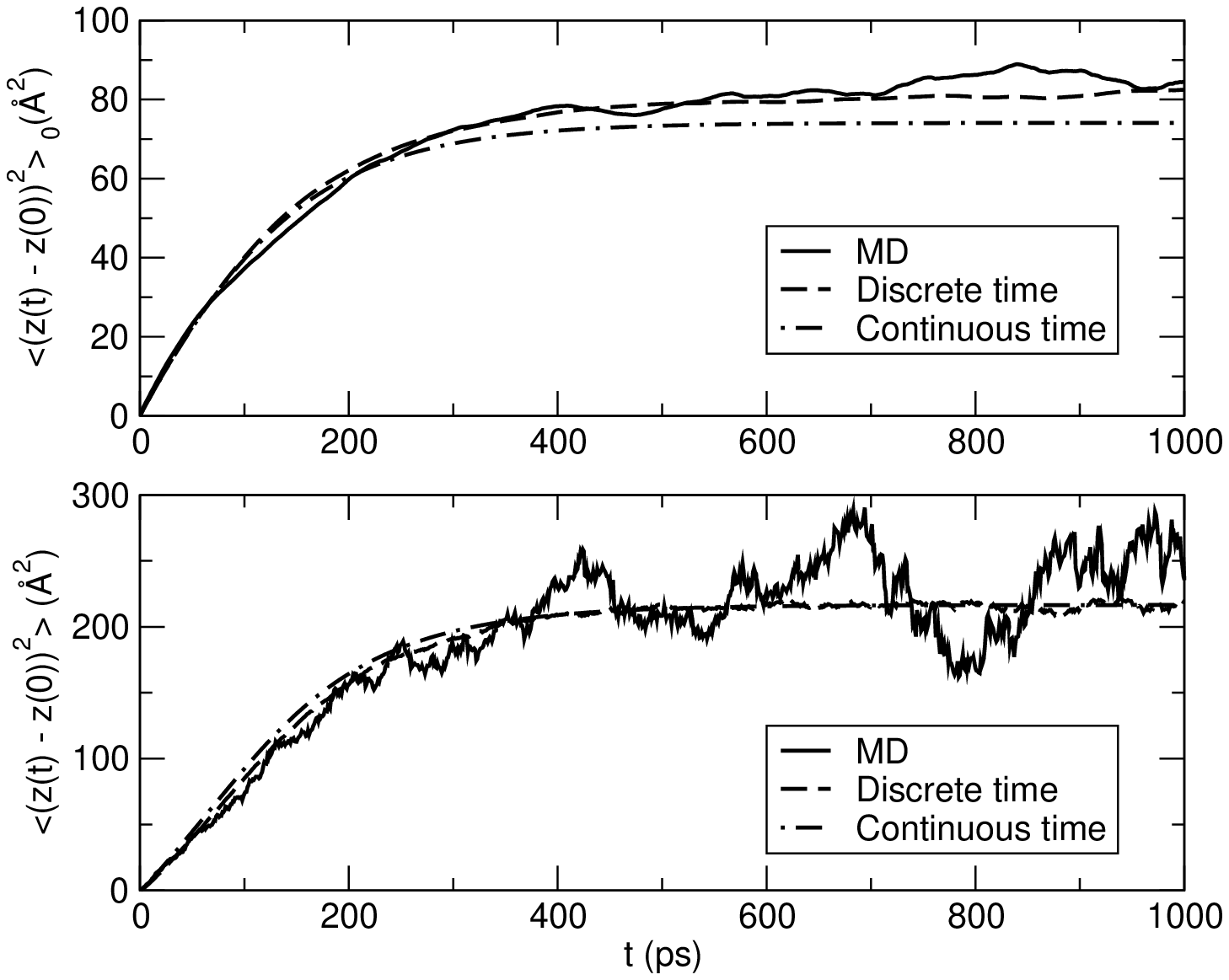} \\
\caption{}
\label{802709JCP4}
\end{figure}

The time-origin averaged MSD of the water molecules inside the 
$28 \AA$ nanotube 
is shown in Figure \ref{802709JCP4} (top panel, solid line). 
The initial 
portion of the curve is linear in time, indicating normal mode diffusion.
At longer times, the curve bends downward and finally it saturates to a 
constant value. The diffusion constant, calculated from the initial slope 
of the MSD versus $t$ curve, is 2.2, 2.5 and 2.8 ($\times 10^{-5} cm^{2}/s$)
for the water molecules inside the $14$, $28$ and the $56 \AA$ 
nanotube, respectively. Even though there
is a slight length dependence, the reason for which is not clear,
the value of the diffusion constant is
roughly half of that of bulk TIP3P water,
$5 \times 10^{-5} cm^{2}/s$.
It is interesting to note at this point that the value of the 
diffusion constant obtained from our calculations is in agreement with that
reported by Berezhkovskii {\it et al.} in Ref.\cite{hummer_prl}.
The diffusion constant estimated from the mean inter particle
separation of $a = 2.6 \AA$ (this is exactly what we get from the
position of the first peak of $g(z)$) and the mean hopping time
$\tau = 13$ ps, reported by Berezhkovskii {\it et al.}, is
$2.6 \times 10^{-5} cm^{2}/s$, which is very close to the values obtained from
our simulations. 

A naive interpretation of the downward bending of the MSD versus $t$ curves
at longer times would be a crossover from normal mode diffusion to 
sub-diffusive behavior as $t$ is increased. In particular, it is possible
to find a time window in which the local slope of the MSD versus $t$ curve in a
log-log plot is approximately equal to $0.5$, the value expected for single file
diffusion. However, such an interpretation of the data is not correct. As
discussed in detail in the next section where we present a simple model for 
the dynamics of the water molecules inside the nanotube and compare the
simulation results for the MSD with analytic ones obtained for this model, 
the MSD data shown in Fig. \ref{802709JCP4} are completely consistent with
normal mode diffusion of the water molecules. The fact that each molecule that
contributes to the calculation of the  MSD for time $t$ 
is required to stay inside the  nanotube of finite length for the entire 
observation time $t$ leads to the downward bending and 
eventual saturation of the MSD at long times. 

Figure \ref{802709JCP4} (bottom panel, solid line) shows the MSD 
calculated with a fixed time origin which is taken to be the time at which each
water molecule enters the nanotube. The MSD shown in this plot is,
therefore, the mean-square displacement of a water molecule inside the
nanotube from its point of entry. This set of data is more
noisy than the time-averaged one due to lack of averaging over
time origins. It is, however, clear that this curve saturates at long times
at a value higher than the saturation value of the time-origin averaged MSD.
Also, the behavior of this curve near $t=0$ is different from that of the
time-origin averaged MSD.
All these features of the results for the fixed time origin MSD can also  
be explained from the random walk models discussed 
in the next section. 

Another important quantity characterizing the dynamics of the confined 
water molecules is the survival probability of the water molecules 
inside the nanotubes. 
This probability was calculated using the definition \cite{pamam}
\begin{eqnarray}
Q(t) = \sum_{i=1}^N < \prod_{t_{k}=t_0}^{t_0+t} P_{i}(t_{k}) >,
\label{surv}
\end{eqnarray}
where the summation is over all the water molecules and the angular brackets
denote an average over the time origin $t_0$. 
The survival probability was normalized 
to unity at $t=0$ by dividing the $Q(t)$ of Eq.(\ref{surv}) by its value at $t=0$. 
Configurations saved at every $1$ ps interval were used in the
calculation of this quantity. $Q(t)$ defined 
above is the probability that a water molecule remains inside the nanotube for
all times between $t_0$ and $t_0+t$, averaged over the initial time $t_0$.
It is related to $p(t)$, the distribution of residence times of the water
molecules inside the nanotube, in the following way.
\begin{equation}
Q(t) = \frac{\int_t^{t_m} (t^\prime-t) p(t^\prime) dt^\prime}{\int_0^{t_m}
t^\prime p(t^\prime) dt^\prime}. \label{def1}
\end{equation}
Here $t_m$ is the total time of the simulation which, for all practical purposes,
may be taken to be infinity.
A related quantity, which is also sometimes called the survival probability,
is $S(t)$, the probability that a water molecule that enters the
nanotube at time $t_0$ remains inside the nanotube for all times between $t_0$
and $t_0+t$. This quantity is related to the residence time distribution
$p(t)$ by
\begin{equation}
S(t) = \int_t^{t_m} p(t^\prime) dt^\prime, \label{def2}
\end{equation}
which implies that $p(t) = -dS(t)/dt$. Using this result in Eq.(\ref{def1}), we
get the following relation between the measured survival probability $Q(t)$ and
the quantity $S(t)$.
\begin{equation}
Q(t) = \frac{\int_t^{t_m} S(t^\prime) dt^\prime}{\int_0^{t_m}
S(t^\prime) dt^\prime}. \label{def3}
\end{equation}
If the residence time distribution decays exponentially with time at long times, 
i.e. if $p(t) \sim \exp(-t/\tau)$, then both $Q(t)$ and $S(t)$ would also decay 
exponentially with the same time scale $\tau$ which would be the average
residence time of the water molecules inside the nanotube.

\begin{figure}[htbp]
\includegraphics[height=8.5cm,width=8.5cm,angle=0]{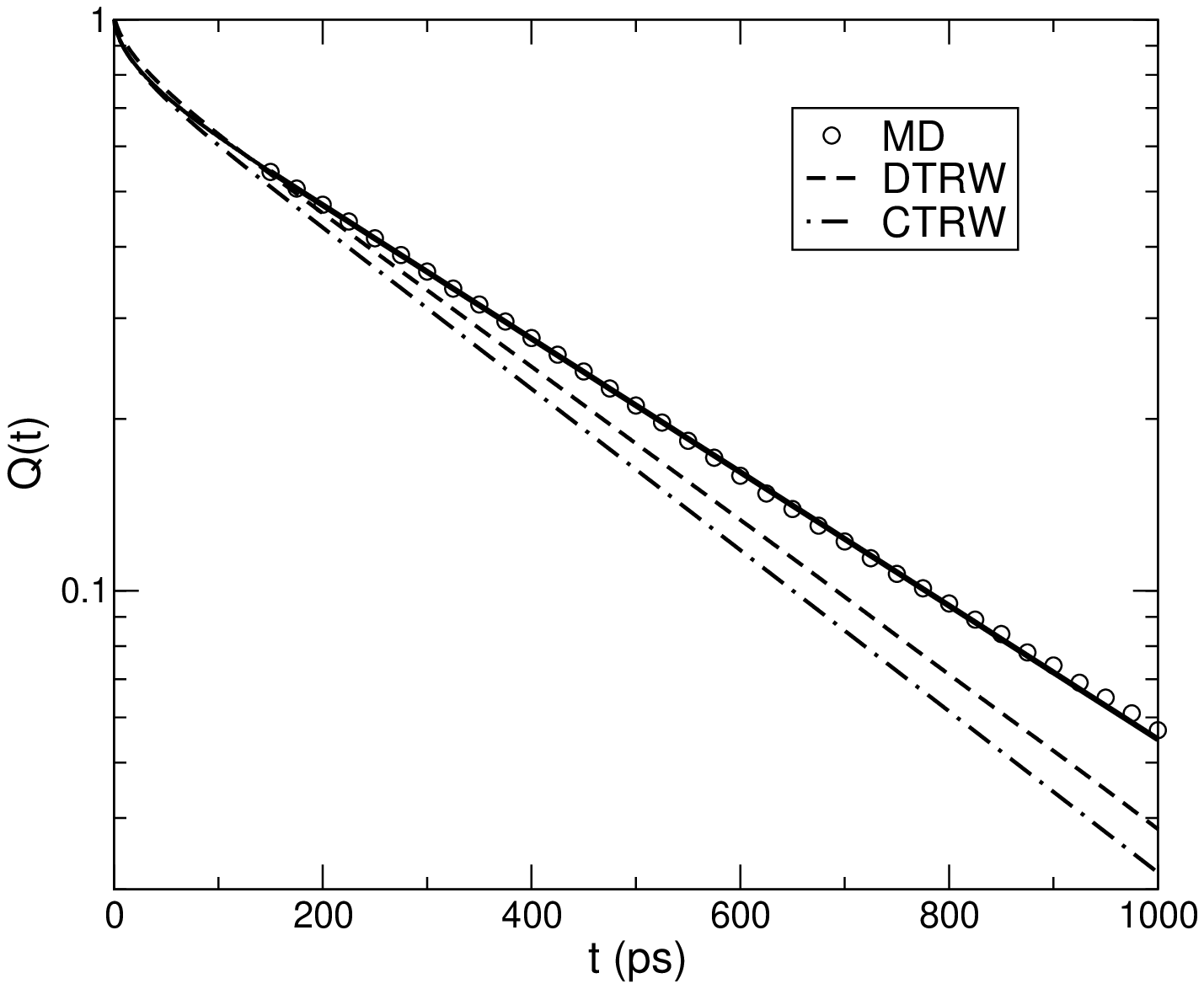} \\
\caption{}
\label{802709JCP5}
\end{figure}

The survival probability $Q(t)$ obtained from MD simulations
for the $28 \AA$ nanotube is shown as a function of time $t$ in 
Fig.\ref{802709JCP5} in a semi-log plot. It is clear from the plot that $Q(t)$
decays exponentially with $t$ at long times. This data is calculated from a 
simulation which ran for $20$ ns.
Results for the $14 \AA$ tube also
show similar behavior. The value of the time constant $\tau$ (the average
residence time) is $73 \pm 0.7$ ps for the $14 \AA$ tube and $371 \pm 5$ ps for the $28 \AA$
tube. These values are roughly 
proportional to $L^2$. The exponential decay
of the survival probability and the quadratic dependence of the time scale
$\tau$ on the length of the nanotube are both consistent with normal mode
diffusion of the water molecules inside the nanotube. This is explained in detail
in the next section.

\section{Random Walk Models and Comparison with MD Results}

Due to the very high degree of correlation among the positions and displacements 
of the confined water molecules demonstrated in the preceding section, 
it is reasonable to form the following approximate description
of the dynamics of the water molecules inside the nanotube: 
the group of water molecules, each separated from its neighbors by the
inter-particle spacing $a$, 
form a ``solid'' object of size comparable to the length $L$ 
of the nanotube.  This group of strongly correlated
molecules moves back and forth along the tube, randomly kicked  
by water molecules from the bath if it happens to be
near an end of the tube. If one of the molecules in this group goes out of the
tube, another one enters the tube within a short time, keeping the total number
of molecules inside the tube roughly constant in time. 
Hence the motion of this group of molecules can be
described as a one-dimensional random walk of a single ``effective'' particle
on a line segment of length $L$. Since we consider the dynamics of the
water molecules only during the time they spend inside the nanotube (a molecule
that goes out of the nanotube becomes a part of the bath and is not
considered in our calculations), the one-dimensional random walk has  absorbing 
boundaries at the two end of the tube ($z=0$ and $z=L$). 
The time $t$ is a continuous variable here. Hence we call this model a 
continuous time random walk (CTRW). All the 
results pertaining to this model can be calculated analytically. We 
have also considered a 
discrete time random walk (DTRW) model (see below) 
for which results were obtained from
Monte Carlo (MC) simulations. 

The diffusion equation for the quantity $P(z,t|z_{0},0)$, 
the probability of finding the particle at $z$ at time $t$, given that the
particle was at $z_{0}$ at the initial time $t = 0$, is given by,
\begin{eqnarray}
\frac{\partial P}{\partial t} = D \frac{\partial^{2} P}{\partial z^{2}},
\label{diff}
\end{eqnarray}
where $D$ is the diffusion constant. 
The absorbing boundary conditions for the above equation are 
\begin{equation}
P(0,t|z_{0},0) = P(L,t|z_{0},0) = 0, \label{bc}
\end{equation}
whereas the
initial condition is
$P(z,0|z_{0},0) = \delta(z-z_{0})$.
The solution of Eq.(\ref{diff}) with the initial and boundary conditions
specified above is
\begin{eqnarray}
P(z,t|z_{0},0) = \sum_{n=1}^{\infty} \frac{2}{L}
\sin\left(\frac{n \pi z_{0}}{L}\right) \sin \left(\frac{n \pi z}{L}\right) 
e^{-D (\frac{n \pi}{L})^{2} t}\,.
\end{eqnarray}
The time-origin averaged MSD can be analytically calculated from the expression
\begin{eqnarray}
< \Delta z^{2} (t) > = \frac{\int_{0}^{\infty} dt_{0}\int_{0}^{L}dz_{2} \int_{0}^{L} dz_{1}
P(z_{2},(t_{0}+t)|z_{1},t_{0}) (z_{2} - z_{1})^{2} P(z_{1},t_{0}|z_{0},0)}
{\int_{0}^{\infty} dt_{0} \int_{0}^{L} dz_{2} \int_{0}^{L} dz_{1}
P(z_{2},(t_{0}+t)|z_{1},t_{0}) P(z_{1},t_{0}|z_{0},0)}\,.
%\nonumber
\end{eqnarray}
The MSD with fixed time-origin, discussed in the preceding section, is given
by 
\begin{eqnarray}
< (z(t) - z_{0})^{2} > = \frac{\int_{0}^{L}dz (z - z_{0})^{2} P(z,t|z_{0},0)}
{\int_{0}^{L}dz P(z,t|z_{0},0)}\,.
\end{eqnarray}
The quantity $S(t)$ defined in the preceding section, which is the probability 
that the particle does not go out of the tube during the time interval between
$0$ and $t$, is given by
\begin{eqnarray}
S(t) &=& \int_{0}^{L} P(z,t|z_{0},0) dz \nonumber \\
     &=& \sum_{n=1}^{\infty} \left(\frac{2}{n \pi}\right)
\sin\left(\frac{n \pi z_{0}}{L}\right) [ 1 - (-1)^{n}]
e^{-D (\frac{n \pi}{L})^{2} t}\,.
\label{st}
\end{eqnarray}
The survival probability $Q(t)$ measured in our MD simulations may be obtained 
by combining this with Eq.(\ref{def3}).

We have used these analytic expressions to calculate all the 
dynamical quantities measured in the MD simulations. 
The input parameters required for comparing the results 
from the random walk model 
with those obtained from MD simulations are the diffusion constant, $D$
of the water molecules inside the nanotube and $z_{0}$, the position of a
water molecule at time $t=0$, i.e. at the moment it comes inside the nanotube.
We used the value of $D$ estimated from the slope of the 
MSD versus time curve
near the origin. The value of $z_{0}$ used in the calculation is
the mean of the starting coordinate of a water molecules inside
the nanotube, measured from the end through which it enters the tube. In making
a comparison of the analytic results for the CTRW model with those obtained from
MD simulations, one should keep in mind the fact that the positions of the 
water molecules are measured at discrete intervals of 1 ps in the simulations.
It is known from earlier studies of similar problems (see, for example,
Ref.\cite{smcd}) that this discrete sampling can make the results for
dynamical quantities significantly different from those obtained from a
continuous-time description. The difference in the present case arises from the 
fact that events in which a water molecule goes out of the tube and re-enters
it between two successive measurements of its position (which are separated
by 1 ps) are not detected in the discrete sampling. This makes 
values of the MSD and the survival
probability obtained from discrete sampling higher than those calculated for
the continuous time model.

To examine the effects of
discrete sampling, we have used MC simulations to study the dynamics of a 
random walk model in which a particle, initially at $z=z_0$, performs a DTRW in 
which the length of each step is a Gaussian random variable with zero mean
and variance equal to $2D\delta t$, where $D$ is the diffusion constant 
obtained from our MD simulation and $\delta t = 1$ ps, the time between two 
successive measurements of the particle position in the MD simulation. The
time in the DTRW model is defined as $n \delta t$ where $n$ is the number of
steps taken by the particle. The MC simulations were carried out with 
absorbing boundary conditions at $z=0$ and $z=L$, and all dynamical
quantities were measured using the same methods as those in the MD simulation.

The results obtained for the CTRW and DTRW models compare well with 
those obtained from our MD simulations.
A comparison of the results for the MSD is shown in
Fig. \ref{802709JCP4} in which 
the top panel shows the time-origin 
averaged MSD, and the bottom panel shows the MSD with 
fixed time origin. The data obtained from the DTRW model are denoted by 
the dashed curve, whereas the analytic results from the CTRW model are 
denoted by the dash-dotted 
curve. While both the random walk models reproduce all the features of the 
MD data, the agreement is better for the DTRW results, as expected.
As in MD simulations, the MSD saturates at long times 
and the value of the MSD at saturation depends 
on the length $L$ of the tube. For the time-origin averaged MSD, 
the analytic CTRW model predicts that the saturation
occurs when $\sqrt{<\Delta z^{2}(t)>}/L \simeq 0.31$, whereas
the MSD with a fixed time origin 
saturates at $\sqrt{<(z(t)-z_0)^2>}/L \simeq 0.52$. This and other differences
between the MD results for the two MSDs are correctly reproduced by the 
random walk models. Similar agreement between MD and random walk results is
also found for the $14 \AA$ nanotube.

Results for the survival probability $Q(t)$ for the $28 \AA$ nanotube 
are compared in Figure 
\ref{802709JCP5}. As expected, the DTRW results for the survival 
probability are slightly higher than the CTRW results. 
Both of these show exponential decay at long times. 
For the CTRW model, the time scale $\tau = L^{2}/(\pi^{2} D)$, since 
the term corresponding to $n = 1$ in Eq.(\ref{st}) dominates at large $t$.
This gives $\tau = 308$ ps for the $L=28 \AA$ tube. The numerical result
for $\tau$ obtained for the DTRW model is $\tau = 328 \pm 10$ ps. These values
are in fair agreement with the result ($\tau = 371 \pm 5$ ps) of our MD simulation.
The results for $\tau$ for the $L=14 \AA$ nanotube are: 84 ps (CTRW), 
$94 \pm 5$ ps (DTRW) and $73 \pm 0.7$ ps (MD). 
The random walk results for $\tau$ show a slight
departure from the expected $L^2$-dependence because the values of $D$ used for
the two values of $L$ are slightly different. Given the extreme
simplicity of the random walk models, the agreement between the random walk
and MD results for the survival probability may be considered quite
satisfactory. The agreement between the model and simulation results for the 
MSD is better than that for the survival probability. This is probably due to
the fact that the details of the complicated process of 
water molecules entering and exiting
the nanotube, which are completely neglected in the random walk model, affect the
survival probability more than the MSD.

It is clear from the results discussed above that the simple one-dimensional random
walk models provide a good description of all the dynamical features found in 
our MD simulations. From this observation, we conclude that the water molecules
undergo normal mode (Fickian) diffusion inside the nanotubes under the conditions
considered in our MD simulations.

\section{Conclusions}
In summary, we have shown that the water molecules inside a narrow (6,6)
nanotube of diameter $8 \AA$ are spatially ordered with a mean
nearest-neighbor distance of $2.6 \AA$.
To mimic the motion of the spatially ordered chain of water molecules 
inside the nanotube,
we propose a random walk model for the
diffusion of a single particle in a finite $1$D channel. The MSD calculated from
this model agrees well with that obtained from the MD simulations.
This establishes that the water molecules undergo normal mode diffusion
inside these short, open-ended carbon nanotubes even for the
longest observation times. It is argued that the reason behind
the observation that the water molecules undergo normal-mode
diffusion and not sub-diffusion, as the geometric constraints would have
suggested, is the strong correlations due to hydrogen bonding between
neighboring water molecules. 
We also measure the survival probability of the water molecules inside
the nanotube from MD simulations and calculate it
from the random walk model. There is good agreement between
the results from the model and MD simulation for 
the time scale (average residence time) of the exponential decay of the 
survival probability. This time scale
is shown to depend quadratically on the length of the nanotube,
which is another piece of evidence in support of the conclusion that 
the mode of diffusion is normal.

Our results have important implications for the possibility of observing 
single file diffusion in narrow channels of finite size. If the particles inside
the channel interact with one another so strongly that they form a single
strongly correlated cluster, then their mode of diffusion would be normal
(Fickian) instead of single-file even if the particles can not cross one another.
This is probably the reason for observing normal mode diffusion in a recent
MD simulation \cite{Striolo_single} of water molecules in a narrow
carbon nanotube with periodic boundary conditions -- the simulation shows that
the water molecules form a tightly bound cluster inside the nanotube. To observe
single file diffusion in such a system, the parameters 
(interaction strength, temperature, length
of the channel, etc.) must be such that the particles inside the channel form
several clusters. It would be interesting to check, by simulations or
experiments, whether a
crossover from normal mode to single file diffusion occurs in such a system 
as appropriate parameters are changed to go from a situation in which 
the particles form a single cluster inside the channel
to one with several clusters.  

\section{Acknowledgments} 
One of us (CD) would like to thank Satya Majumdar for helpful
discussions. AKS thanks the Deptartment of Science \& Technology, India for financial support.
BM thanks JNCASR for financial support.

\pagebreak
\section{Figure Captions}

Fig. 1. The pair correlation function $g(z)$ for the water molecules
inside the $14$, $28$ and $56 \AA$ nanotubes.
The well separated
peaks signifies a solid like positional ordering of the water molecules inside the
nanotube even at a temperature of $300$ K. The average spacing between
the water molecules is about $2.6 \AA$. \\\\

Fig. 2. In order to probe the effects of the open ends of the nanotube
on the positional ordering of the water molecules we have compared the
positional correlation of only the water molecules residing in the
central $14 \AA$ within the $28 \AA$ nanotube, with those inside the
$14 \AA$ tube. The fact that both the correlation functions are identical
confirms that the degree of positional correlations is same in both the nanotubes. \\\\

Fig. 3. The quantity $<\delta z_{i}(\delta t) \delta z_{j}(\delta t)>$
as a function of the separation between the ``i'' th and
the ``j'' th molecule,
in units of $a_{0}$, the mean inter-particle spacing, whose value
is $2.6 \AA$. The correlation of dispacements at two sites decays
as one increases the separation between the sites. \\\\

Fig. 4. The time-origin averaged MSD (top panel) and the
fixed time-origin MSD (bottom panel) for the water molecules
inside the $28 \AA$ nanotube. The solid curve is data from MD,
dashed curve denotes data from DTRW and the dash dotted curve
is data from CTRW. \\\\

Fig. 5. The survival probability, $Q(t)$, for the water molecules inside the
$28 \AA$ nanotube.
The long time data obtained from MD is denoted by circles. This data has been
fitted to the function $exp(-t/\tau)$, the solid line is the fit for which
$\tau = 371 \pm 5$ ps. The dashed line indicates
$Q(t)$ obtained from DTRW, whereas the
dash-dotted curve is result from CTRW.

\pagebreak
\newpage
\begin{table}
\label{lenjonesparm}
\centering
\begin{tabular}{|c|c|c|}
\hline
$Species$&$\epsilon (kcal/mol)$&$\sigma (\AA)$ \\ \hline
$Carbon(CA)-Carbon(CA)$&$0.086$&$3.4$ \\ \hline
$Carbon(CA)-Oxygen(OW)$&$0.11$&$3.27$ \\ \hline
\end{tabular}
\caption{The Lennard Jones interaction parameters between various atomic species.}
\end{table}

\clearpage

\pagebreak

\pagebreak

\pagebreak

\pagebreak

\end{document}